# Millimeter-Wave for Unmanned Aerial Vehicles Networks: Enabling Multi-Beam Multi-Stream Communications

Yiming Huo, Member, IEEE, and Xiaodai Dong, Senior Member, IEEE

{ymhuo@uvic.ca, xdong@ece.uvic.ca}

Department of Electrical and Computer Engineering University of Victoria, BC V8P 5C2, Canada

*Abstract*–With the fifth-generation (5G) mobile networks being actively standardized and deployed, many new vehicular communications technologies are developed to support and enrich various application scenarios. Unmanned aerial vehicle (UAV) enabled communications emerges as one of many promising solutions of constructing the next-generation highly reconfigurable and mobile networks. In this article, we first investigate and envision the challenges of future UAV applications from the network, system, and hardware design perspectives, and then present a UAV aerial base station (ABS) prototype which works at millimeter-wave (mmWave) bands and enable multi-beam multi-stream communications. In terms of the field trial tests of the first UAV-ABS of its kind in the world, multi-giga-bit-per-second data rate of uplink and downlink is verified with good stability and reliability against mildly challenging weather conditions.

## I. INTRODUCTION

Over the past decade, due to many technological catalysts such as the breakthroughs in artificial intelligence (AI), wireless communications, computer vision, mechanical and industrial design, space and unmanned aerial vehicle (UAV)/drone technologies advance dramatically, which has made these technologies more affordable and accessible to civilian and commercial applications. DJI Innovations, Parrot, SpaceX, Loon LLC are representative corporations associated with the rapidly growing market in recent years. In the meantime, this phenomenal change has enabled more feasible and cost-effective way of using UAVs to better serve human society and activities in our daily life. Besides executing cargo transportation, entertaining, security patrol, agricultural applications, etc., UAVs will profoundly alter the way we think about wireless communications particularly the cellular broadband communications that has been dominated by the terrestrial communications for the last several decades.

As summarized in [1], UAV-aided wireless communications are categorized into three major types: first, UAV-aided ubiquitous coverage where existing communication infrastructure are assisted by UAVs for seamless wireless coverage; second, UAV-aided relaying for providing wireless connectivity to distant users or user groups without direct and/or reliable communication links; third, UAV-aided information dissemination and data collection where delay-tolerant information and data are collected from distributed wireless devices, e.g., internet of things (IoT) applications. More specifically, UAV-aided ubiquitous coverage and UAV-aided relaying encompass the application scenario of deploying UAVs to form aerial base station (ABS) enabled communications networks with high mobility. For example, it is valuable and practical in providing temporary/recovery service when terrestrial base stations (BSs) malfunction, or fast handling an abrupt increase of cellular service demands that appear in a specific region.

In terms of standardization progress, a study item on enhanced long-term evolution (LTE) support for connected UAVs was started in Release 15, by the third Generation Partnership Project (3GPP) in March 2017 [2], and completed in December 2017, with LTE UAV field test results documented and analyzed [3], for sub-6GHz bands. Both academia and industry have demonstrated prototype designs [4], field test results, and UAV capable cellular networks designs. For example, Google Loon project enabled emergency LTE service recovery to Puerto Rico after the Hurricane Maria disaster in 2017; Qualcomm and AT&T test UAVs on commercial LTE networks for accelerating wide-scale deployment; Verizon has been testing a 200-pound gas-powered drone in New Jersey, for providing a 4G LTE signal throughout a one mile range. The existing UAV-ABS prototypes and field trials for mobile communication purpose have been demonstrated in 4G LTE bands, with satisfactory coverage but limited RF bandwidths and thus data rate.

Millimeter-wave (mmWave) bands offer much wider available bandwidths and hence have attracted significant attention for use in 5G terrestrial communications. For aerial communications, reduced antenna dimension at the mmWave bands has a particular advantage: smaller and lighter physical footprint on payload [5]. This article explores mmWave bands for UAV-ABS and presents a distributed phase array architecture that supports multi-beam, multi-stream and multi-user (MU) communications, verified by proof of concept (PoC) field tests. In particular, this article covers the following aspects:
1) UAV dispatching and communications enabling strategy in UAV-ABS fast response application scenario;
2) Distributed mmWave antenna array beamforming modules (BFM) for UAV-ABS based on channel characteristics, phased-array antennas simulations, and application scenarios;
3) Multi-beam multi-stream communications system for UAV-ABS: from concept, circuit and system, to overall hardware design and implementation;
4) MmWave multi-beam multi-stream UAV-ABS proof-of-concept field trial with test results presented and analyzed.

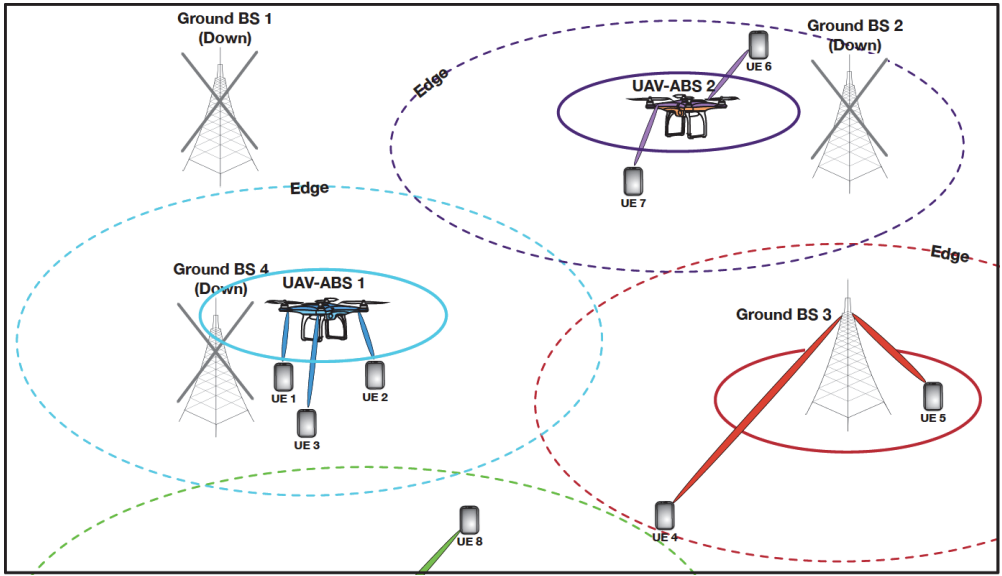

Fig. 1. Application scenario of UAV-ABS in fast response to communication service recovery.

## II. UAV-ABS Fast Response Application Scenario

As illustrated in Fig. 1, in a specific region when terrestrial communications are unavailable or of inadequate performance, multiple UAV-ABSs are dispatched to enable the 5G/B5G services for ground users (GUs). For example, UEs 1-3, and UEs 6-7 are supported by UAV-ABS 1 and 2, respectively, when ground BS services are interrupted out of emergency. All three previously mentioned UAV use cases fall into the depicted application scenario.

Reasonable path planning and trajectory optimization [6] are mandatory for UAV enabled communication networks considering the typical constraint of the limited on-board power and thus the flight time [7]. Furthermore, the energy efficiency of the aerial equipment is determined to a large extent by the wireless system design, which is more critical for a UAV system with limited payload and on-board energy. Therefore, it is desired that a future UAV-ABS system is able to serve multiple ground users with large bandwidths at mmWave bands, with minimum co-channel or adjacent channel interference.

On the other hand, the critical factors impacting terrestrial mmWave communications, such as shadowing effects and large propagation loss, can be overcome by UAV-ABS with beamforming, beam-tracking, high mobility and self-organizing networking capabilities. As envisioned, unlike the ground BS, the UAV-ABS is able to timely adjust location, keep reasonable distance and flying attitude with, and maintain suitable direction towards ground user(s). The UAV-GU distance is determined by various factors such as communication quality, user experience feedbacks (tolerance to noise), safety standards, etc.

On the other aspect of UAV-ABS dispatching and communication establishment strategy, a feasible strategy of UAV-ABSs serving GUs is depicted in Fig. 2. For the initialization step, the GPS location information of the no-service or poor-service areas is sent to the assigned UAV-ABSs by the backbone network which has evaluated the emergency severity and calculated the solutions.

After being dispatched to the locations where ground BS services malfunction, via wireless signaling and advanced artificial intelligence enabled computer vision and sensing technology on board the UAV-ABS, ground users can be searched and accurately localized. The UAV-ABS then identifies and approaches the ground users, but with a suitable distance kept when enabling communication channels. Flying attitude, status, and velocity are timely adjusted in terms of the channel status information (CSI), communication quality, and user experience feedbacks. For mmWave communication using beamforming technology, with the assistance of both deep learning (DL) enabled computer vision and real-time CSI, optimal beam alignment can be maintained. In the meantime, the backhauling can be simultaneously conducted by the UAV-ABS using another mmWave band.

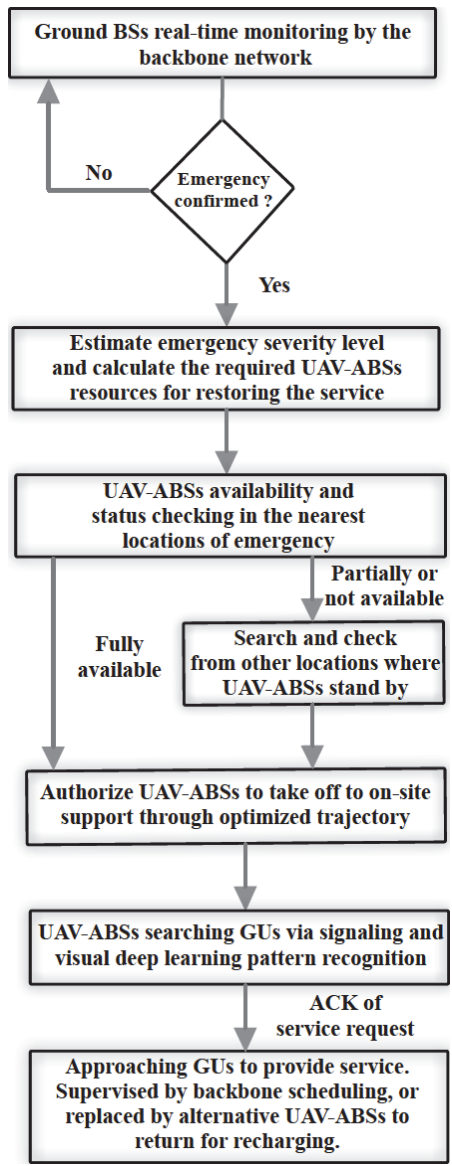

Fig. 2. Strategy of dispatching UAV-ABSs providing emergency service recovery.

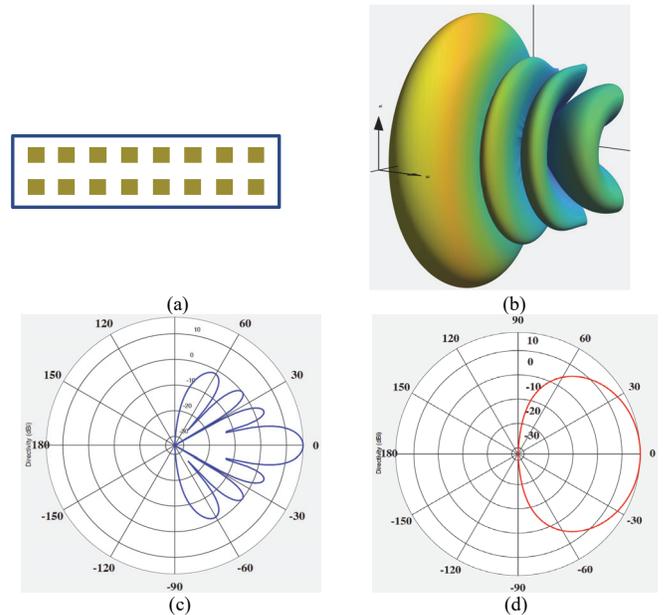

Fig. 3. (a) 2×8 phased array, and (b) 3D radiation pattern, (c) azimuth plane radiation pattern, (d) elevation plane radiation pattern, when steering angles are set to 0.

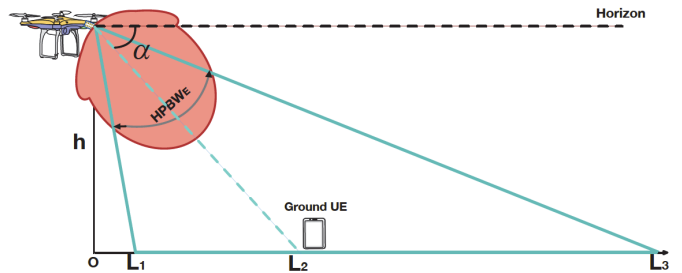

Fig. 4. Side view of typical SU application scenario of UAV-ABS.

### III. UAV-ABS SYSTEM ARCHITECTURE AND IMPLEMENTATION

#### A. UAV-ABS Channel Characteristics

According to mmWave terrestrial channel model study and field trials [8]-[10], mmWave channels are characterized by high propagation loss, strong shadowing effects and blockage by human bodies. Phased array enabled 3D beamforming is one of many appealing solutions [11]. Although there is limited prior work on mmWave Air-to-Ground (A2G) and Air-to-Air (A2A) channel modeling, different from terrestrial communication channels, A2G and A2A mmWave have smaller path loss exponent and lighter small-scale fading [12]. Moreover, 3D beamforming enables higher probability to achieve line of sight (LoS) in comparison with terrestrial communications. Steering mmWave beams can help mitigate the co-channel interference in the MU mode and enhance the communication secrecy. Therefore, 3D beamforming is suitable and critical for the UAV-ABS to conduct ground cellular service, and A2G/A2A backhauling/front-hauling.

One unique challenge of UAV-ABS channels unlike its geographically fixed terrestrial BS counterpart is that UAV-ABS is highly mobile and the movement behavior, such as drifting, rotation, tilting, vibration, etc., is determined by the aerodynamic interactions between the environment and UAV mechanical and navigation systems. The corresponding dynamic channel properties need further study by theoretical modelling and field measurements.

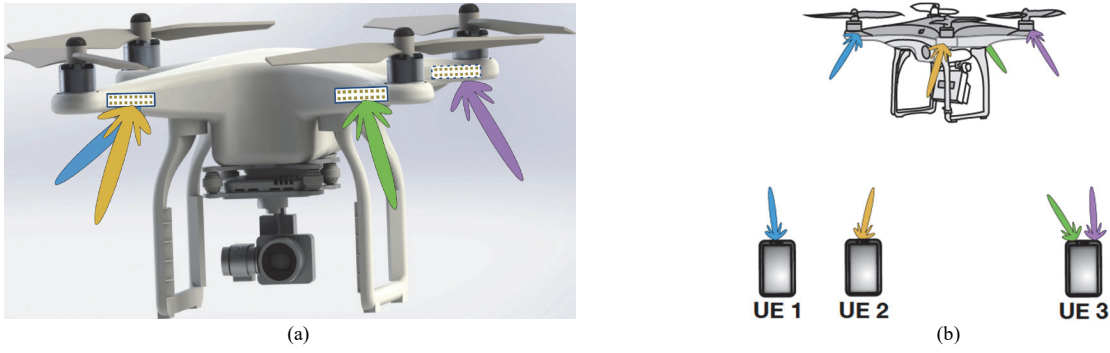

Fig. 5. (a) UAV-ABS product concept design, and (b) its enabling multiple beams and streams communications with multiple users at mmWave bands.

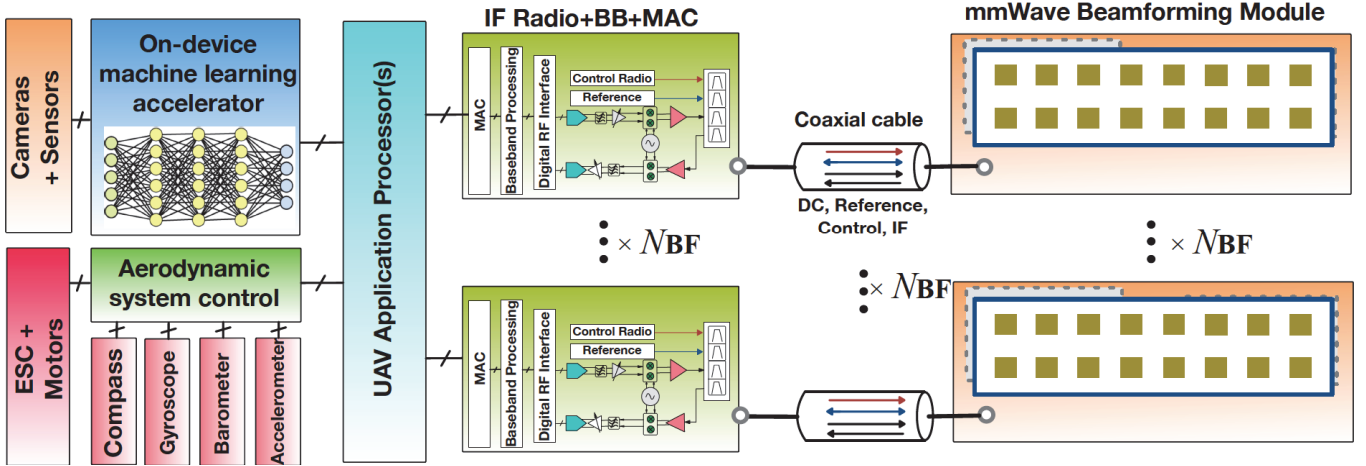

Fig. 6. Proposed system and circuit design of machine learning assisted mmWave mulit-beam multi-stream UAV-ABS.

For A2G/A2A channels, enabling mmWave 3D beamforming capable UAV needs to fulfill several requirements as follows: first, the uniform planar array (UPA) is mandatory to provide both azimuth and elevation dimensions; second, the broad steering degree of either azimuth or elevation angles is desired to enhance the coverage and facilitate the establishment of the optimal communication channels. Both SU and MU modes should be supported to cope with more diversified application scenarios, and more importantly, reduce the total system resource consumption.

### B. Phased Array Beamforming Module for UAV-ABS

In our design case, a uniform rectangular based phased array with a dimension of 2 × 8 antenna elements is employed to realize 3D beamforming. This is mainly because, considering the mobility, movement direction and quasi-stationary characteristics of the UAV-ABS, the elevation radiation pattern of the main lobe should have enough wide beamwidth to cover the targeted ground UEs while the azimuth radiation pattern of main lobe needs to obtain narrow beamwidth to increase the directivity and facilitate better beam-steering resolution and less cross-beam interference. Fig. 3(a) illustrates the 3D radiation pattern when the steering angles are set to (0, 0) where side lobes are minimized with a maximum array gain of 16.46 dBi from simulations. As illustrated in Fig. 3(b), the half-power beamwidth (HPBW) of the array in the azimuth plane, $HPBW_A$, is 10.5°, and the side lobe level (SLL) is -13.1 dB. Fig. 3(c) depicts the radiation pattern in the elevation plane, and the $HPBW_E$ is around 60°.

Such $HPBW_A$ and $HPBW_E$ can facilitate the following typical SU application scenario of the UAV-ABS as shown in Fig. 4. Assume UAV-ABS is working at height $h$, and the downtilt angle is $\alpha$, the horizontal coverage distance $L_3$-$L_1$ is calculated as

$$L_3 - L_1 = h \times [\tan\left(90° - \alpha + \tfrac{1}{2} \cdot HPBW_E\right) - \tan\left(90° - \alpha - \tfrac{1}{2} \cdot HPBW_E\right)] \quad (1)$$

where $\alpha$ is smaller than 90°, and $h$ is determined by several constraints, namely path loss, blockage, safety, noise level (at ground UE end), federal aviation administration (FAA) regulations, etc. If $h$ is 10 m, $L_3$-$L_1$ is 25.7 m when $\alpha$ is set to 50°. A larger $\alpha$ results in smaller coverage, but reduced path loss.

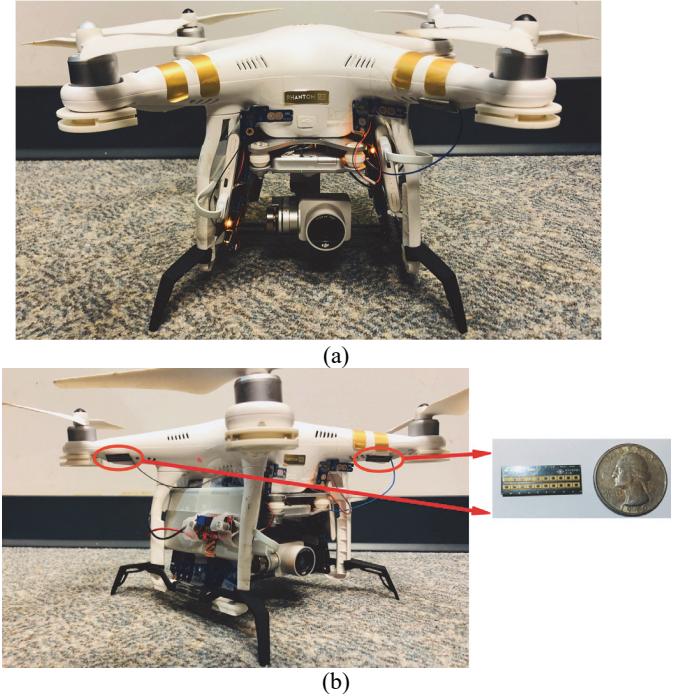

Fig. 7. Overall design and implementation of UAV-ABS system (a) front view, and (b) side view with slim form-factor beamforming module.

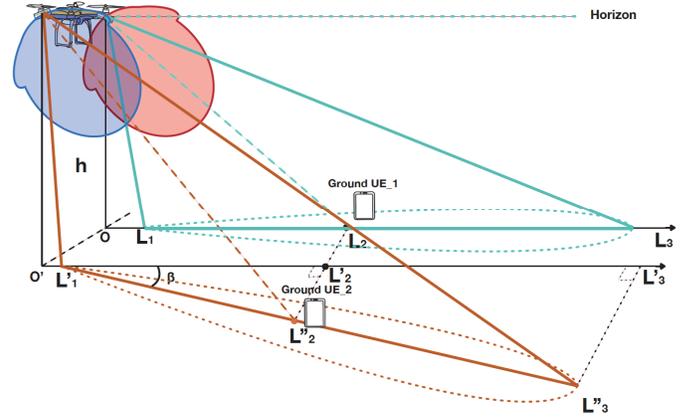

Fig. 8. 3D view of MU (two users) application scenario of UAV-ABS.

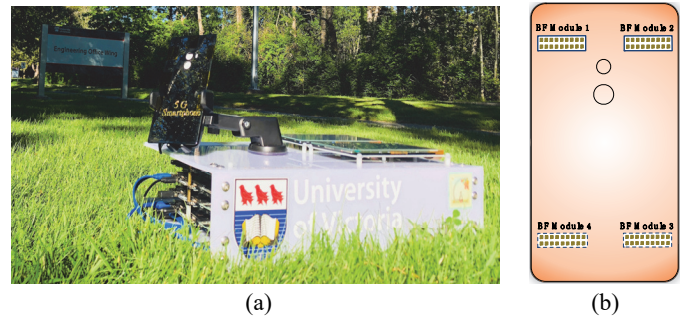

Fig. 9. (a) DPA-MIMO proof-of-concept (PoC) system, and (b) mmWave beamforming module placement demonstration on smartphone UE prototype.

### C. Multi-beam and Multi-stream System Design

Employing phased array for 3D beamforming creates highly directional radiation patterns and communications channels. To cover omnidirectionally positioned GUs and overcome the instability caused by the aerodynamic behavior of the UAV, it is necessary to distribute the phased array BFM to different parts of the UAV body while connect and control them in a central manner to coordinate beamforming for spatial multiplexing of multiple data streams to single or multiple users. As illustrated in Fig. 5 (a), multiple (= 4) phased arrays are arranged and placed in a distributed manner on the arms of DJI Phantom 3, with enough separation (e.g, edge-to-edge distance >> 2•$\lambda_0$ where $\lambda_0$ is free-space wavelength) maintained among different phased arrays to maximize the spatial multiplexing gain and minimize the cross-beam and co-channel interference. In addition, the interference can be significantly reduced when different carrier frequencies are chosen.

As further illustrated in Fig. 5 (b), the application scenario of communication with three ground UEs is supported. Each beam of one specific color represents one independent data stream carried. These UEs can be distributed phased array based MIMO (DPA-MIMO) enabled UE devices [13]. For example, UE 3 that accommodates two beamforming modules is able to scale up the transmission speed almost twice theoretically. Due to the narrow HPBW$_A$ and sufficiently large isolation among phased arrays on the UAV, high data rate can be obtained.

As depicted in Fig. 6, a UAV-ABS system employs multiple mmWave beamforming modules (compact designs) that contain both phased-array antennas and RF front-end components such as power amplifiers (PAs), low-noise amplifiers (LNAs), phase shifters, frequency synthesizers, mixers, multiplexers, etc. $N_{BF}$ mmWave BFMs are judiciously embedded over the UAV body and connected through
coaxial cables to intermediate frequency (IF) radios, baseband processing unit (BBPU), MAC protocols processors (MPPs), UAV application processors (APs), respectively. Power supply (DC), reference signals, control signals, and IF signals are conveyed through coaxial cables. Such distributed phased arrays architecture can facilitate many advantages for the UAV enabled mmWave communications, e.g., high beam-domain spatial multiplexing gain, low interference, better thermal performance, high reconfigurability, etc.

Furthermore, besides the mmWave communications part, the UAV-ABS application processors (APs) are connected with two other major function blocks, i.e., aerodynamic system control which is connected to compass, gyroscope, barometer, accelerator, electric speed control (ESC) with motors, etc.; the second function block is mainly for synthetic intelligent visual and sensing, which are realized by high dynamic range (HDR) cameras, sensors, and on-device machine learning accelerators. As discussed in Fig. 2, the UAV-ABS needs to localize and approach the GUs through wireless signaling and machine learning assisted sensing (optical and/or sound). It is desired that the machine learning models can be locally operated on a UAV-ABS instead of solely relying on cloud computing empowered machine learning engines, so that lower latency and better reliability can be achieved. Fortunately, more and more commercial ready on-device AI chipset solutions are making this highly possible [14].

Finally, a MU application scenario example is illustrated in Fig. 8 in which two ground UEs are separated with a distance $L_2$-$L''_2$, and $\beta$ is the intersection angle formed by the projection of two BFMs' main lobes in the elevation pattern, on the ground. Straight line $L_3$-$L_1$ and $L''_3$-$L'_1$ stand for the projection of the maximum elevation gain of two BFMs on the ground, respectively. With a suitable angle $\beta$ obtained via steering the two beams, cross-beam interference can be significantly mitigated.

*D. System Implementation*

The entire system hardware implementation is subject to the extra payload the UAV is able to lift as well as limited on-board power supply, therefore the overall hardware design has been optimized and minimized. As illustrated in Fig. 7, two 60 GHz beamforming modules are fixed onto two arms of the DJI Phantom 3. The BFM supports three WiGig channels which work at center frequencies of 58.32 GHz, 60.48 GHz, and 62.64 GHz, respectively. The main logic board (MLB) accommodates IF radio and baseband stage circuitry which realizes WiGig MAC, modem, BF algorithm, array control, etc. High-efficiency DC-DC converters are used to transform the output of independent power supply from 5 V to 19.5 V. The dimensions of BFM are $25 \times 9 \times 2$ mm$^3$, weighing only 1 gram. The antenna array is designed with patch antenna elements that are spaced by half of free-space wavelength (at 62.5 GHz), plus 4 dummy antenna elements on both sides to enhance the performance. The overall extra payload is maintained at around 1.2 pound including USB 3.0 flash drives used to provide test files.

## IV. FIELD TRIAL TEST

*A. MmWave Massive MIMO Capable Ground User Equipment*

The ground UE prototype is implemented using the DPA-MIMO architecture and design method [13], [15], as shown by the DPA-MIMO proof-of-concept (PoC) in Fig. 9(a). Moreover, the placement of BFMs on UE housing is illustrated in Fig. 9(b) where totally four BFMs are distributed and positioned on the UE, with an edge-to-edge separation larger than 1.5 free-space wavelength for adjacent BFMs. During the field trial test for the SU application scenario, we enable two BFMs on the top frame; for later MU test, we enable one BFM on each UE.

*B. Test Scenarios and Test Results*

Field trial tests are conducted in a local area shown in the Google's 3D map of Fig. 10(a), with trees and bushes around as shown in Fig. 10(b). There are two typical scenarios where the UAV-ABS provides mmWave beamforming enabled giga-bits-per-second (Gbps) link service, i.e., SU and MU mode.

In the SU scenario, the UAV-ABS enables two beams to transmit two data streams while hovering in the sky at a height and with some specific distance away from the ground UEs that accommodate two mmWave BF modules. A peak downlink speed of (PDLS) of 2240 Mbps (measured using large compressed documents to transfer from UAV to the UE) is achieved. When the UAV-ABS hovers around 41 m away from the ground UE ($h$ is 35 m, and $d_0$ is 22 m), the communication link is still very stable.

It is worth mentioning that, compared to terrestrial communications when the signal strength and connection become weak and unstable as the distance goes over 30 m, A2G communication demonstrates advantages in the field trial test, which complies with the A2G channels characteristics holding smaller path loss exponent and lighter small-scale fading.

In the MU scenario tests, two ground UEs with single BFM enabled on each are deployed with separation distance, $d_1$, varied to receive individual beam ($d_1$ and $d_0$ are perpendicular, and $d_1$ is set to 6 m, and 2 m, respectively). UE 1 and UE 2 achieve an aggregated PDLS of 2168 Mbps (1096 Mbps and 1072 Mbps for UE 1 and UE 2, respectively). In another test scenario, when UE 2 is placed at Location 2 with $d_2$ ($d_0$ and $d_2$ are parallel) set to 10 m and the other parameters are the same as before, the aggregated PDLS is slightly decreased to 2136 Mbps. Therefore, a larger intersection angle, $\beta$, due to wider separation of ground users can result in reduced BFMs' maximum gain, which is in accordance with the phased array simulation results conducted for beamforming design.

It should be pointed out that, the quasi-stationary status of UAV in which small movement and slow horizontal drifting happens (e.g. from Location I to Location II, or Location I to Location III, as shown in Fig. 10 (a)) due to limited localization accuracy and abrupt atmospheric pressure change, does not affect the performance of the wireless communications in our field trial tests. It is also confirmed that, even when the wind speed is 6-7 m/s reported on ground (higher wind speed with higher altitude [15]), the communication quality is stable, which can be explained by, first, the aerodynamic system can facilitate necessary system-level robustness and reliability against environmental variance; second, the beamforming and wireless design are appropriate and assist solving the environment-induced impairments. For example, a 10.5 degree of $HPBW_A$ in this design can tolerate a large horizontal shift compared to the distance. Moreover, the beam misalignment can be quickly fixed by beam tracking and re-alignment from the algorithms. In near future, as more industrial level UAVs will appear in the market, UAV-ABSs can provide even better comprehensive performance and immunity to adverse weather and environment.

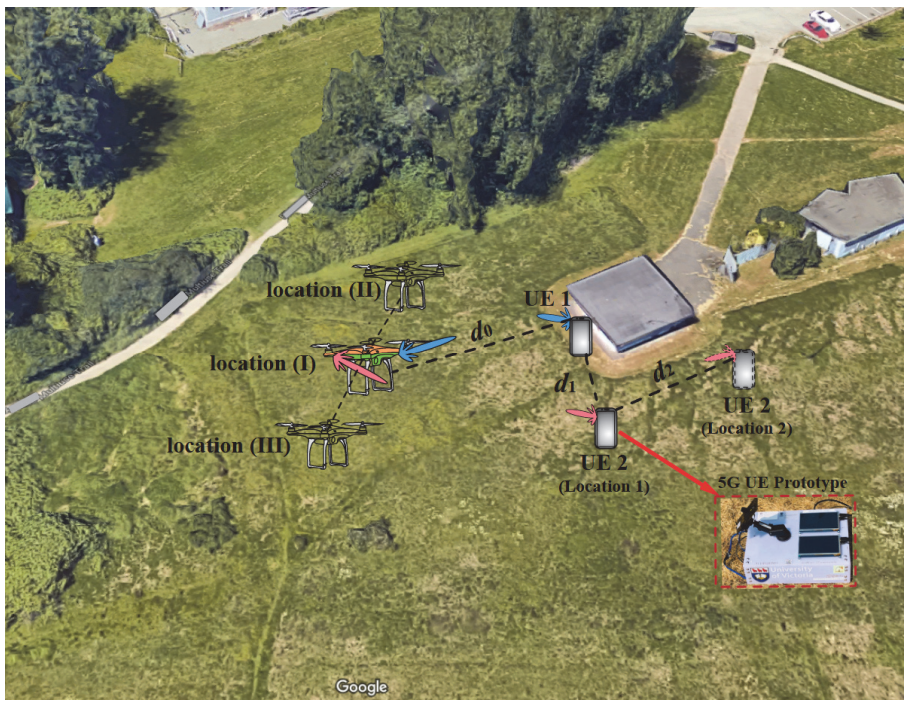 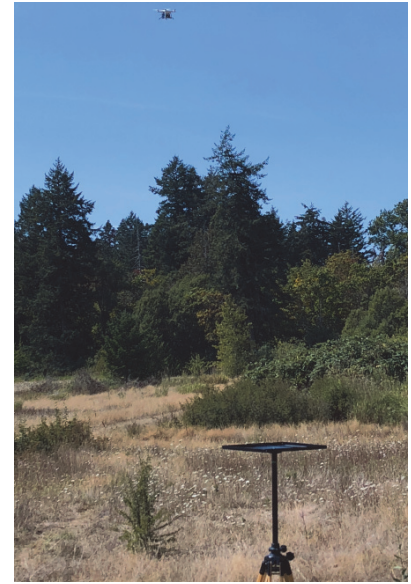

Fig. 10. (a) Field trial environment with various application scenarios tested and verified, and (b) UAV-ABS hovers and provides communication service.

*C. Other Design Concerns*

User experience is another major concern of UAV-ABS system design. For example, we have measured that, the sound level when UAV hovers at 1 m away from the user is as high as 88 dB (85 dB or stronger sound can cause permanent hearing damage) while the sound level decreases to 66 dB when the UAV is separated from the user by 10 m. Considering that acoustic wave is a type of mechanical waves with different attenuation characteristics than radio waves, more study and research need to be conducted on modeling UAV acoustic noise, and reducing it from several aspects such as noise-cancelling aerodynamic designs, user-friendly path planning, etc.

## V. CONCLUSIONS

This article mainly deals with the future mmWave communication capable UAV-ABS system, through investigation and analysis of application scenarios, deployment scheme and strategy, enabling technologies including massive MIMO plus machine learning, and design and verification of a prototyping system. We believe that, UAV-ABS requires a comprehensive system design methodology involving aerodynamics, wireless communications, and artificial intelligence. To the best of our knowledge, this is the first UAV based ABS proof-of-concept (PoC) design working at mmWave bands and facilitating multi-beam multi-stream communications. Field trial tests have presented promising results for both single-user and multi-user scenarios to deliver multi-gigabit-per-second wireless data links, which shows the extraordinary potential of UAV-ABS systems for future 5G/B5G communications networks.

ACKNOWLEDGMENTS

The authors would like to acknowledge Natural Sciences and Engineering Research Council of Canada, and Franklin Lu for help on the UAV field tests.